\documentclass{PoS}
\usepackage{mcite}
\usepackage{subfigure}

\title{Correlations of electrons from heavy flavor decay in p+p, d+Au and Au+Au collisions}

\ShortTitle{Correlations of electrons from heavy flavor decay in p+p, d+Au and Au+Au collisions}

\author{\speaker{Anne Sickles, for the PHENIX Collaboration}%
       Brookhaven National Laboratory\\
       E-mail: \email{anne@bnl.gov}}

\abstract{
In relativistic heavy ion collisions heavy flavor probes are crucial to understand the interactions between partons and the produced hot nuclear matter. Measurements in p+p collisions provide information about how the heavy quarks are produced and fragment and in d+Au collisions are sensitive to possible effects from cold nuclear matter. Azimuthal correlation measurements involving heavy flavor probes are complementary to single particle spectra measurements and provide additional information about production and interactions of heavy quarks. Measurements of electrons with heavy flavor decay with other hadrons from the event can provide information about how the heavy quark interacts with the produced matter and can be compared to similar measurements from light hadron correlations. Correlations between electrons from heavy flavor decay with muons, also from heavy flavor decay, can provide further information about heavy flavor production and cold nuclear matter effects in d+Au collisions with a very clean signal. We present PHENIX results for electron-hadron correlations in p+p and Au+Au collisions and electron-muon correlations in p+p and d+Au collisions and discuss the implications of these measurements.

}

\FullConference{XVIII International Workshop on Deep-Inelastic Scattering and Related Subjects\\
		 April 19 -23, 2010\\
		 Convitto della Calza, Firenze, Italy}

\begin{document}


Angular correlations particles from the decay and fragmentation of heavy quarks is of great interest
in hadronic and nuclear collisions.  In $p$+$p$ collisions these measurements can aid in the
understanding of the heavy flavor production mechanisms.  In small nuclear systems, such as
$d$+Au collisions these correlations can be sensitive to cold nuclear matter and gluon
saturation effects.  In heavy ion collisions, these measurements are crucial to understanding
heavy quark energy loss in hot nuclear matter.  A substantial suppression of electrons from
the decay of $D$ and $B$ mesons with respect to expectations from $p$+$p$ collisions has been observed at 
mid-rapidity~\cite{ppg066}.  This was unexpected if the hard partons moving through the hot nuclear
matter lost energy primarily via gluon bremsstrahlung because of the charm and bottom quarks' large
mass~\cite{dead_cone}.
However interpretation of correlation measurements  
requires measurements in $p$+$p$ collisions to establish a reliable baseline.  The
PHENIX experiment at RHIC has measured azimuthal correlations of electrons from heavy flavor decay
and charged hadrons in $p$+$p$ and Au+Au collisions and the correlations between electrons
and muons at forward rapidity, where both leptons are from heavy flavor decay in $p$+$p$ and
$d$+Au collisions at $\sqrt{s_{NN}}=$200GeV.

In PHENIX heavy flavor is measured via leptons from the semileptonic decay of D and B mesons.
PHENIX has excellent electron identification capabilities at mid-rapidity $|\eta|<$0.35.  
At more forward and backward rapidities, 1.2$<|\eta|<$2.4, muons are measured.  With the
planned installation of silicon vertex detectors at both mid-rapidity (2010) and forward rapidity
(2011) further separation between charm and bottom decay products can be made, but currently
they are combined into a single sample.  
For electrons at mid-rapidity PHENIX has measured the fraction of electrons from bottom decay
by looking at the invariant mass of $e-K$ pairs and found it to be in agreement with theoretical 
calculations~\cite{ppg094}.  In 
the electron transverse momentum, $p_{T,e}$,
 range of interest here, $p_{T,e}<$4.5 GeV/c 
$\approx$10\%-50\% of the electrons are from B decay; the fraction increases with increasing 
$p_{T,e}$.

\begin{figure}
\centering
\subfigure{
\includegraphics[width=0.45\textwidth]{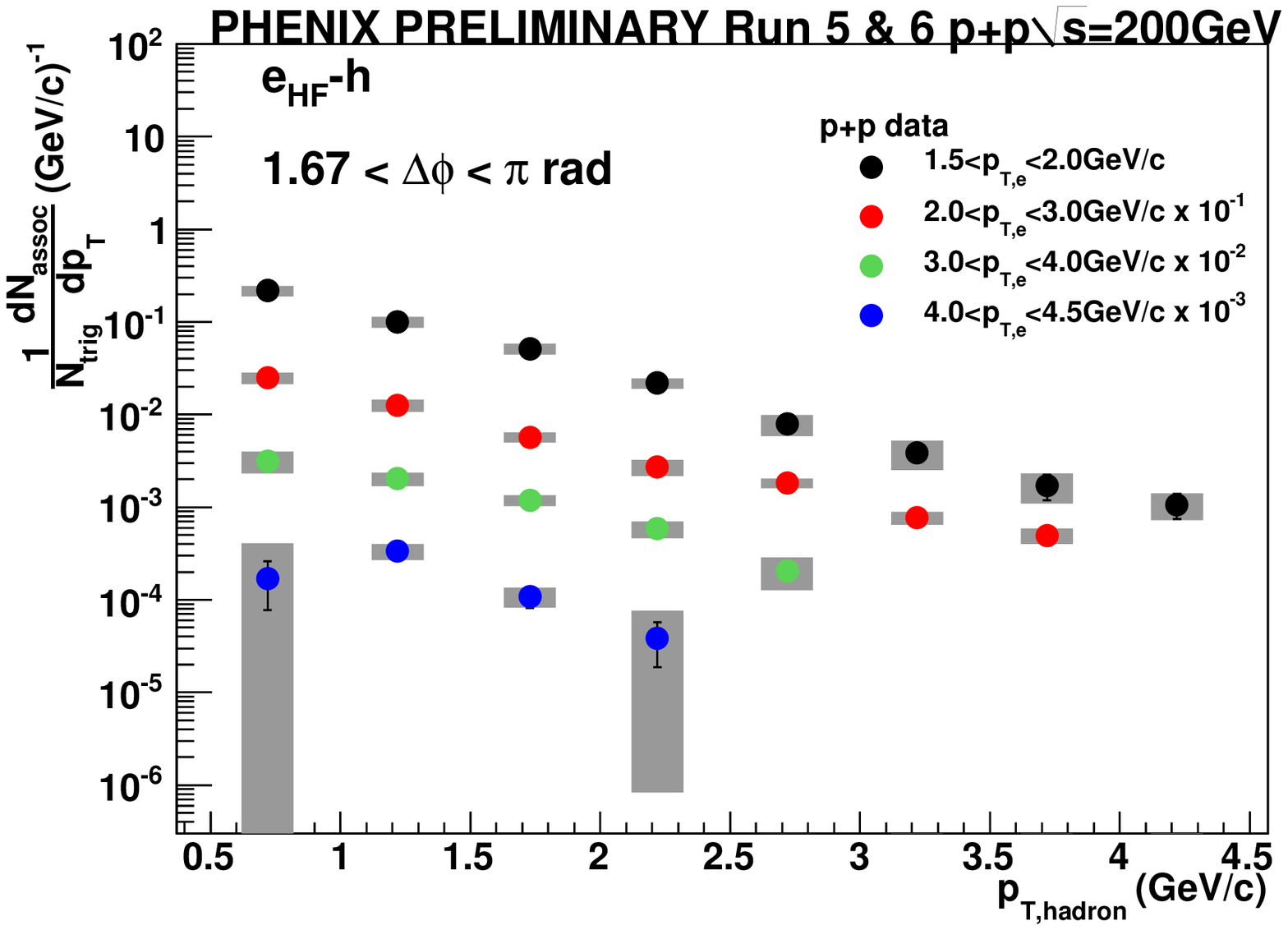}
\label{pp_yield}}
\subfigure{
\includegraphics[width=0.45\textwidth]{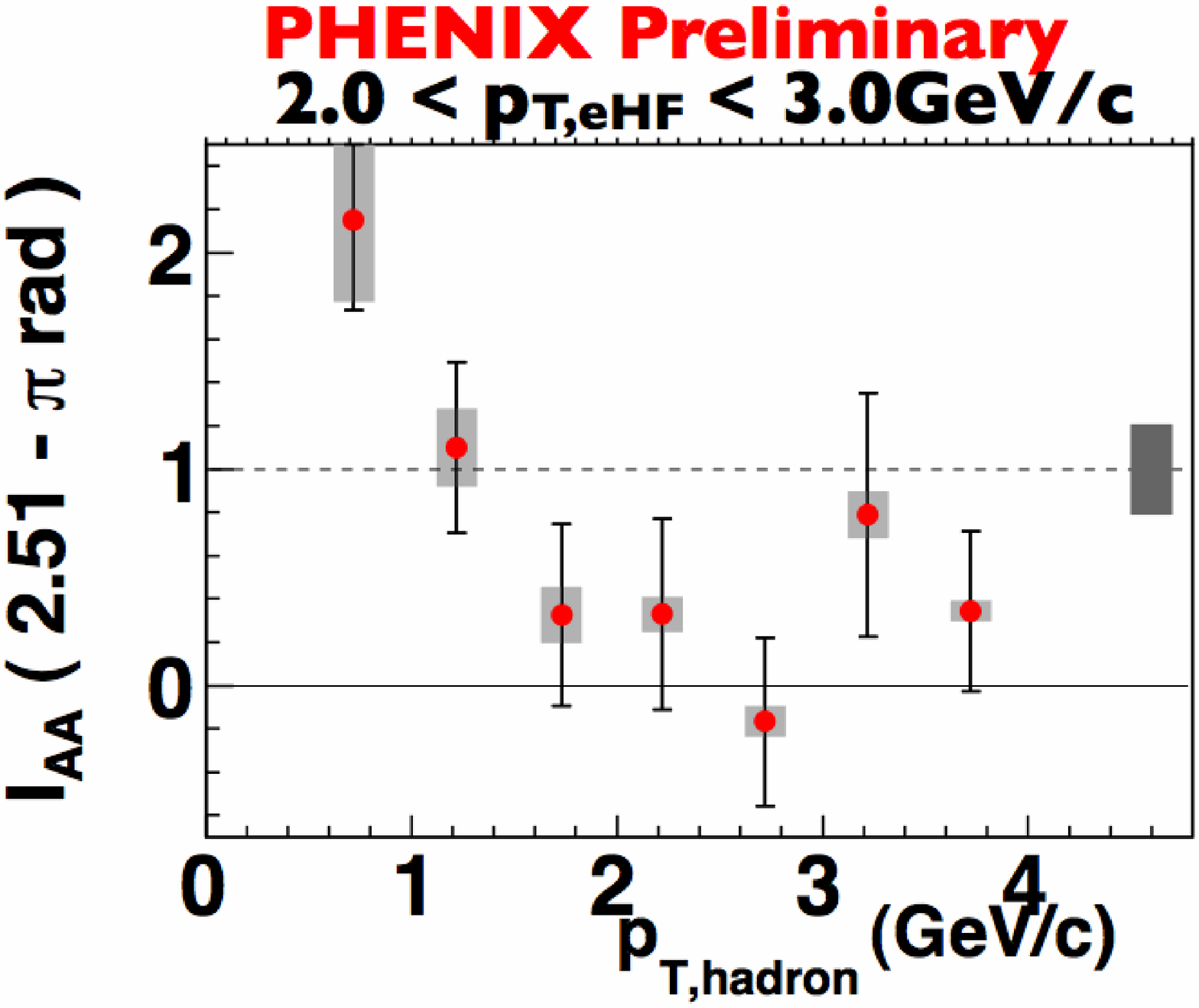}
\label{iaa}}
\caption{Left: Away side conditional yields for $e_{HF}-h$ correlations
for four electron $p_T$ bins and eight hadron $p_T$ bins. 
Right: $I_{AA}$ on the away side for 2.0$<p_{T,e}<$3.0 GeV/c electron triggers as a function of $p_{T,h}$.}
\label{away_fig}
\end{figure}

Azimuthal angle correlations at mid-rapidity have been used extensively at RHIC to complement
single particle yield measurements.  Correlations between trigger particles, in this case electrons
from heavy meson decay, and charged hadrons are measured.  Pairs correlated only through
event wise correlations are subtracted~\cite{abs}.  Remaining pairs at this momentum range are correlated through
being products of the same hard parton-parton scattering.
Measured electron-hadron ($e_{inc}-h$) correlations are a weighted average of the correlations
from electrons from heavy meson decay  and those from background sources:
\begin{equation}
Y_{e_{inc}-h}(\Delta\phi) = \frac{N_{e_{HF}}Y_{e_{HF}-h} + N_{e_{bkg}}Y_{e_{bkg}-h}}
{N_{e_{HF}} + N_{e_{bkg}}}
\end{equation}
$Y$ is the correlated yield of hadrons as a function of $\Delta\phi$ per electron.
  Here the background
electrons are dominately from $\pi^0$ and $\eta$ Dalitz decays and photon conversions in air
and detector material.  The photons which convert are also primarily from $\pi^0$ and $\eta$ decay.
The background contribution can be determined from other measurements (including, for example, the $\pi^0$
$p_T$ spectrum), so the ratio $R_{HF} = \frac{N_{e_{HF}}}{N_{e_{bkg}}}$ can be 
determined~\cite{ppg065,ppg066}.  $Y_{e_{HF}}(\Delta\phi)$ can be expressed as (leaving off the
$\Delta\phi$ dependence of all $Y$ terms):
\begin{equation}
Y_{e_{HF}-h} = \frac{(R_{HF}+1)Y_{e_{inc}-h} - Y_{e_{bkg}-h}}{R_{HF}}.
\label{sub_eq}
\end{equation}
The remaining unknowns
 are the correlations of the electrons from background sources with hadrons, $Y_{e_{bkg}-h}$.
These are determined from measured photon-hadron correlations (like the background electrons,
inclusive photons are dominantly from $\pi^0$ and $\eta$ decay at these $p_T$s); simulations
are used to account parent meson distributions in the decay electron and decay $\gamma$ samples~\cite{wwnd09}.
Fig.~\ref{pp_yield} shows the azimuthally
integrated away side yield ($\Delta\phi\approx\pi$), as a function of the $p_T$ of the associated hadron for four
electron $p_T$ selections.


Fig.~\ref{iaa} shows the away side yield in Au+Au divided by the yield in $p$+$p$ collisions,
$I_{AA}$.  The result suggests some away side suppression of the parton opposing the
leading heavy quark traverses the hot nuclear matter and is qualitatively consistent with the
$I_{AA}$ measured in hadron-hadron correlations~\cite{ppg083}.

\begin{figure}[t]
\centering
\includegraphics[width=\textwidth]{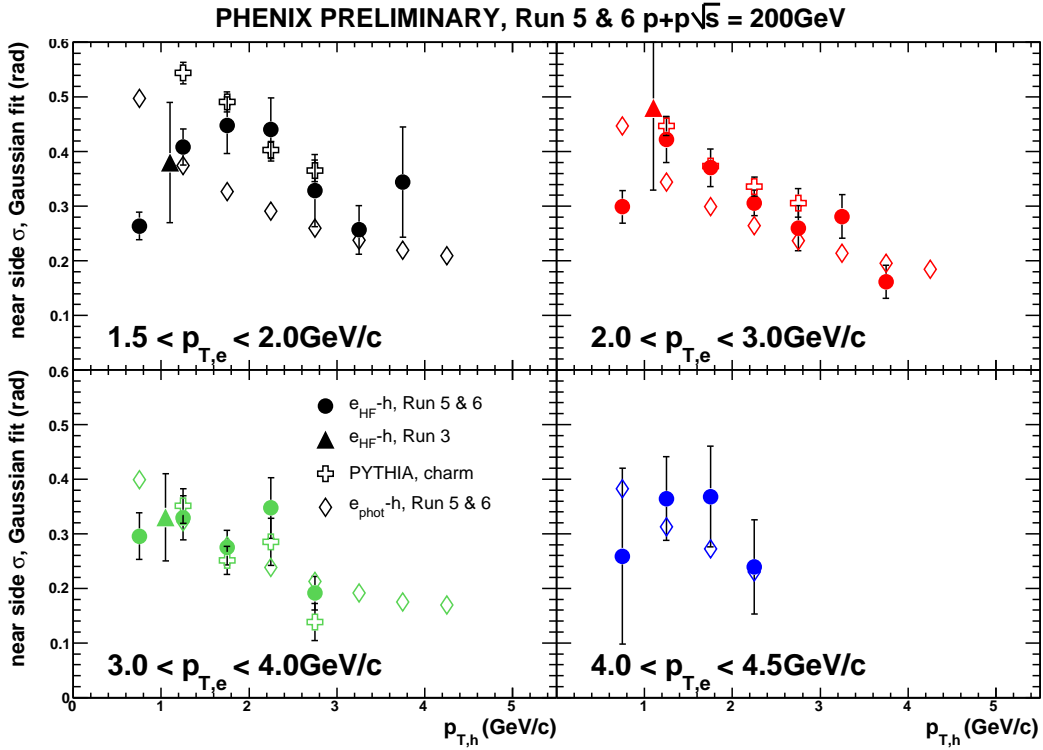}
\caption{Gaussian widths as a function of $p_{T,h}$ for four electron $p_T$ bins.  Circles
show the $e_{HF}-h$ widths and open diamonds show $e_{phot}-h$ widths.  Pluses show
$e_{HF}-h$ widths from PYTHIA and the triangles show an earlier PHENIX measurement
of $e_{HF}-h$ widths with lower statistics and 1.0$<p_T<$4.0GeV/c.}
\label{widths}
\end{figure}

In heavy quark fragmentation most of the quark momentum is carried by the heavy 
meson~\cite{cleoc,*bellec,alephb,*opalb,*sldb}, thus
the near side $e_{HF}-h$ is expected to be dominated by pairs in which both particles
come from the heavy meson decay.  Due to the decay kinematics, we then expect 
the Gaussian widths of the near side to be wider for $e_{HF}-h$ than $e_{inc}-h$.  The widths
from both categories of electrons are shown in Fig.~\ref{widths}.  The measured 
near side widths for
$e_{HF}-h$ are wider than $e_{phot}-h$ and are in agreement with widths from charm production
in PYTHIA~\cite{pythia}.

\begin{figure}
\centering
\subfigure{
\includegraphics[width=0.45\textwidth]{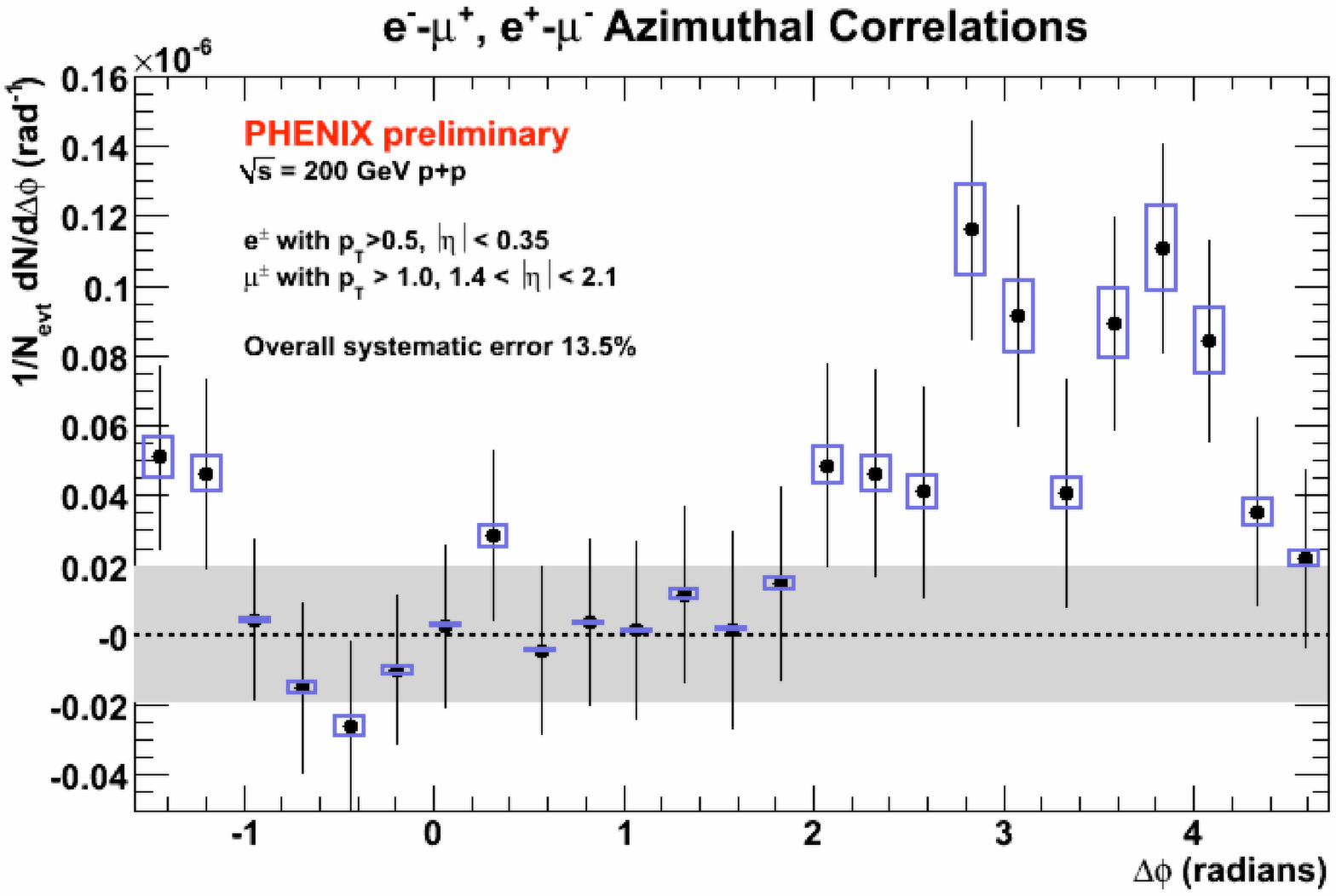}
\label{emu_pp} }
\subfigure{
\includegraphics[width=0.45\textwidth]{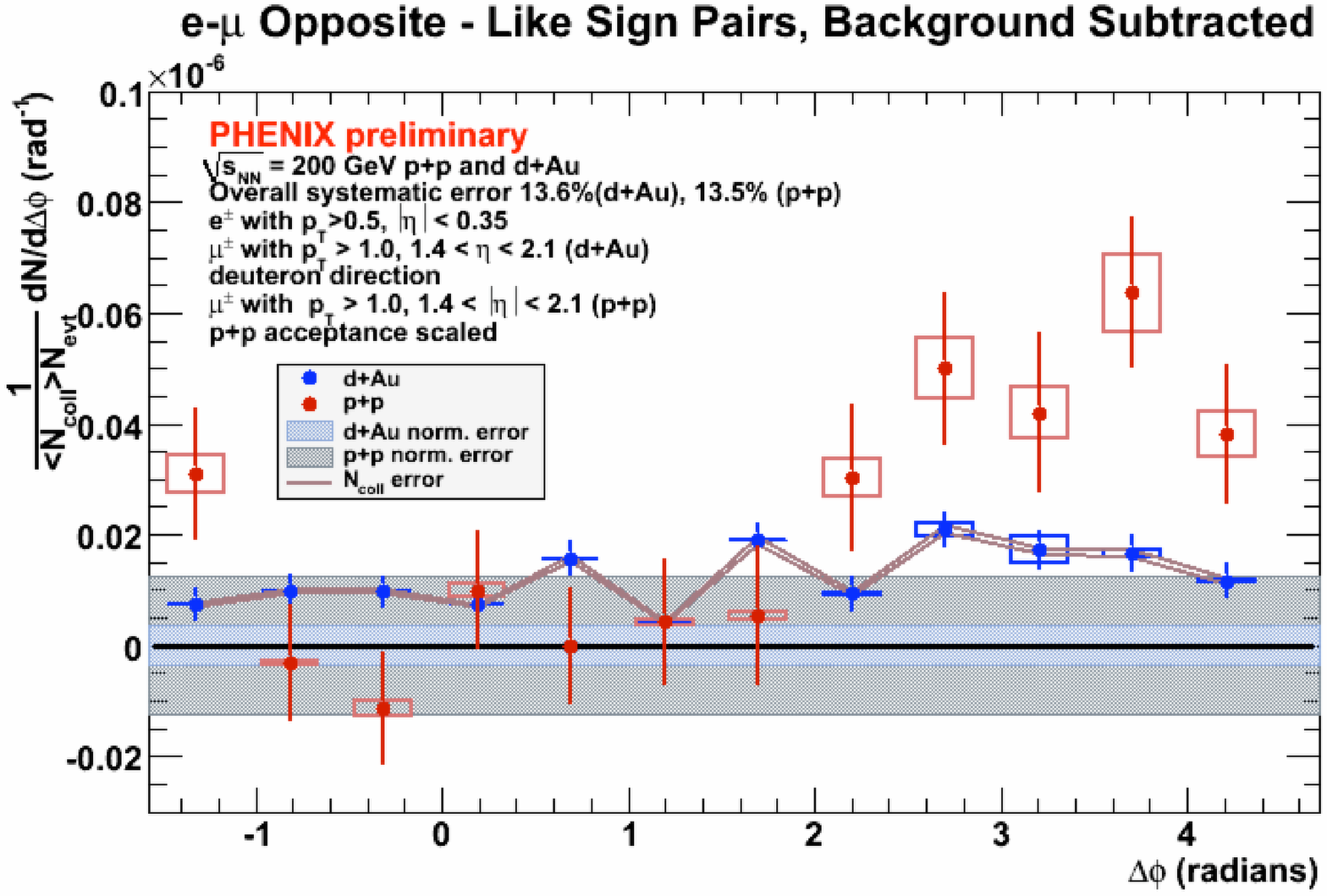}
\label{emu_dAu} }
\caption{Azimuthal distributions of opposite sign electron-muon pairs per event after background
subtraction in $p$+$p$ (left) and $d$+Au (right) collisions.  The $d$+Au distribution has
been scaled down by the mean number of binary collisions in the event sample in order to 
be compared to $p$+$p$ collisions.}
\label{emu}
\end{figure}

Electron-muon correlations were proposed long ago as a clean $c\bar{c}$ signal in heavy ion
collisions~\cite{gavin96}.  
One advantage to measuring electron-muon correlations is that both particles are from the 
decay of heavy flavor.  The correlated charm signal produces electrons and muons of opposite 
sign from each other, thus most backgrounds such as jets and combinatoric pairs
are removed by subtracting like sign pairs.  Remaining backgrounds are due to leptons
not from charm decay.  Fig.~\ref{emu_pp} shows the final electron-muon azimuthal correlations in
$p$+$p$ collisions\cite{tatia_qm09}.  Electrons have $p_T>$0.5 GeV/c and $|\eta|<$0.35
and muons have $p_{T}>$1.0 GeV/c and 1.4$<|\eta|<$2.1.
The near side peak is missing because the pairs are from back-to-back $c\bar{c}$
pairs; any near side correlation would not extend over the large rapidity difference between
the two leptons.  The same measurement in $d$+Au collisions 
with the muon going in the deuteron going direction (1.4$<\eta<$2.1)  is shown in Fig.~\ref{emu_dAu}.  Obviously,
a $d$+Au event has a larger average pair yield than a $p$+$p$ event.  Thus the $d$+Au  data has
been scaled down by the number of average number binary nucleon-nucleon collisions in the event
sample, $N_{coll}$.  This is appropriate because the creation of the $c\bar{c}$ pairs is a hard
process which should scale with $N_{coll}$.  At small $\Delta\phi$ the scaled $d$+Au and $p$+$p$
collisions are consistent with each other, but at $\Delta\phi\approx\pi$ the $d$+Au 
pair yields are significantly suppressed with respect to $p$+$p$ collisions.

With the muon at forward rapidity (deuteron going direction) these correlations are sensitive to the
low $x$ region of the Au nucleus.
study of the $x$ region of the Au nucleus probed in this kinematic region shows some
sensitivity to $x\leq$0.01 where saturation effects could possibly be significant~\cite{kharzeev}.

In summary, PHENIX has measured the azimuthal correlations of leptons from heavy flavor decay
in three collision systems.  These measurements are sensitive to the details of heavy 
flavor production and its modification in nuclear collisions.  A significant suppression of
rapidity separated electron-muon pairs is observed in $d$+Au collisions.  Further studies are
ongoing.    In Au+Au collisions we see
suppression of away side associated hadrons similar to those from hadron triggered correlations.  
While the uncertainties in these measurements are large, measurements will be improved in the
near future from upgrades to the PHENIX detector, including silicon vertex detectors at
mid-rapidity and forward rapidity and increased statistics.  

\bibliography{sickles_dis2010}
\bibliographystyle{iopart-num}

\end{document}